\documentstyle[12pt,epsfig]{article}
\begin{document}

\begin{center}
\vskip 0.4cm

{\Large {\bf {High sensitivity GEM experiment on 2$\beta $ decay of $^{76}$%
Ge }}}
\end{center}

\vskip 0.7cm

\begin{center}
{\bf Yu.G.~Zdesenko}\footnote{%
Corresponding author. Address: Institute for Nuclear Research, Prospekt
Nauki 47, MSP 03680 Kiev, Ukraine; fax: 380 44 265 4463; phone: 380 44 265
2210; e-mail: zdesenko@kinr.kiev.ua}{\bf , O.A. Ponkratenko, V.I. Tretyak}

\vspace{0.5cm}

{\it Institute for Nuclear Research, MSP 03680 Kiev, Ukraine}

\hspace{1.0in}

{\bf Abstract}
\end{center}

\noindent
The GEM project is designed for the next generation $2\beta $ decay
experiments with $^{76}$Ge. One ton of ''naked'' HP Ge detectors (natural at
the first GEM-I phase and enriched in $^{76}$Ge to 86\% at the second GEM-II
stage) are operating in super-high purity liquid nitrogen contained in the
Cu vacuum cryostat (sphere $\oslash $5 m). The latest is placed in the water
shield $\oslash $11$\times $11 m. Monte Carlo simulation evidently shows
that sensitivity of the experiment (in terms of the $T_{1/2}$ limit for $%
0\nu 2\beta $ decay) is $\approx $1$0^{27}$ yr with natural HP Ge crystals
and $\approx $1$0^{28}$ yr with enriched ones. These bounds corresponds to
the restrictions on the neutrino mass $m_\nu \leq 0.05$ eV and $m_\nu \leq
0.015$ eV with natural and enriched detectors, respectively. Besides, the
GEM-I set up could advance the current best limits on the existence of
neutralinos -- as dark matter candidates -- by three order of magnitudes,
and at the same time would be able to identify unambiguously the dark matter
signal by detection of its seasonal modulation.

\hspace{1.0in}

\vskip 0.5cm




\section{Introduction}

At present, when neutrino physics has undergone a revolution (see \cite
{Zub98} and refs. therein), the $2\beta $ decay search plays even more
important role in modern physics\footnote{%
The neutrinoless (0$\nu )$ double $\beta $ decay is forbidden in the
Standard Model (SM) since it violates lepton number ($L$) conservation.
However many extensions of the SM incorporate $L$ violating interactions and
thus could lead to 0$\nu $2$\beta $ decay. In that sense 0$\nu $2$\beta $
decay has a great conceptual importance due to the strong statement obtained
in a gauge theory of the weak interaction that a non-vanishing 0$\nu $2$%
\beta $ decay rate, independently on which mechanism induces it, requires
neutrinos to be massive Majorana particles \cite{Val82}.} than several years
ago \cite{Moe94,Tre95,Fae98,Klap98,Suh98,Vog00}. Indeed, the solar neutrino
problem \cite{Kir99}, the measured deficit of the atmospheric muon neutrinos
flux \cite{SK98} and the result of the LSND accelerator experiment \cite
{LSND00} could be explained by means of the neutrino oscillations, requiring
in turn nonzero neutrino masses ($m_\nu $). However, oscillation experiments
are sensitive to neutrino mass difference, while only measured neutrinoless
(0$\nu $) double $\beta $ decay rate can give the absolute scale of the
effective Majorana neutrino mass\footnote{%
Obviously, its accuracy depends on the uncertainties of the nuclear matrix
elements calculation.}, and hence provide a crucial test of neutrino mass
models \cite{Klap00,Bil99}. Therefore, the 0$\nu $2$\beta $ decay is
considered as a powerful test of new physical effects beyond the SM. The
absence of this process yields strong restrictions on $m_\nu $, lepton
violation constants and other parameters of the manifold SM extensions,
which allow to narrow a wide choice of the theoretical models and to touch
the multi-TeV energy range competitive to the accelerator experiments \cite
{Fae98,Klap98,Suh98,Vog00,Klap00}.

Despite the numerous efforts to detect 0$\nu $2$\beta $ decay, this process
still remains unobserved \cite{Moe94,Tre95}. The highest half-life limits
were set in direct experiments with several nuclides: $T_{1/2}^{0\nu }\geq
10^{22}$ yr for $^{82}$Se \cite{Se82}, $^{100}$Mo \cite{Mo100}; $%
T_{1/2}^{0\nu }\geq 10^{23}$ yr for $^{116}$Cd \cite{Cd2000}, $^{128}$Te, $%
^{130}$Te \cite{Te130}, $^{136}$Xe \cite{Xe136}; and $T_{1/2}^{0\nu }\geq
10^{25}$ yr for $^{76}$Ge \cite{Ge76,IGEX}. These results have already
brought the most stringent restrictions on{\ the values of the} Majorana
neutrino mass {$m_\nu \leq $ (0.5--5.0) eV, right-handed admixture in the
weak interaction }$\eta $ $\approx ${\ 1$0^{-7}$, $\lambda $} $\approx $ 1$%
0^{-5}${, the neutrino-Majoron coupling constant $g_M$ }$\approx ${\ 1$%
0^{-4} $, and the }$R$-parity\footnote{$R $-parity is defined as $R_p$ = ($%
-1 $)$^{3B+L+2S}$, where $B$, $L$ and $S$ are the baryon and lepton numbers,
and the spin, respectively.} violating parameter of minimal SUSY standard
model $\varepsilon ${\ }$\approx ${\ 1$0^{-4}$}. {However, on the basis of
current status of astroparticle physics it is very desirable to improve the
present level of sensitivity} by one-two orders of magnitude \cite
{Zub98,Klap98,Vog00}.

Many projects were proposed during a past few years with regard to these
goals, however most of them require strong efforts and long time to prove
their feasibility (see next section). To this effect, in the present paper
we suggest the GEM project of the high sensitivity 2$\beta $ decay
experiment with $^{76}$Ge, those accomplishment seems to be realistic.
Before enter upon the project itself (section 3), the sensitivity
limitations and current status of the 2$\beta $ decay studies, as well as
requirements to the future projects are considered briefly in section 2.

\section{Sensitivity limitation, present status and future of 2$\beta $
decay studies}

There are two different classes of $2\beta $ decay experiments: with
''passive'' source, which can be simply placed as foil between two
detectors, and with ''active'' source, where detector containing $2\beta $
candidate nuclei serves as source and detector simultaneously \cite
{Moe94,Tre95}. If neutrinoless $2\beta $ decay occurs in the ''active'' or
''passive'' source, the sharp peak at the $Q_{\beta \beta }$ value would be
observed in the electron sum energy spectrum of the detector(s). The width
of this peak is determined by the detector energy resolution. The
sensitivity of the set up for $2\beta $ decay study can be expressed in
terms of a lower half-life limit as following \cite{Moe94,Tre95}: $%
T_{1/2}\sim \eta \cdot \delta \sqrt{(m\cdot t)/(R\cdot Bg)}.$ Here $\eta $
is the detection efficiency; $\delta $ the abundance or enrichment of
candidate nuclei contained in the detector; $t$ the measuring time; $m$ the
total mass of the ''active'' or ''passive'' source; $R$ the energy
resolution ({\it FWHM}) of the detector; and $Bg$ the background rate in the
energy region of the $0\nu 2\beta $ decay peak (expressed, for example, in
counts/yr$\cdot $keV$\cdot $kg).

First of all, it is clear from this equation that efficiency and enrichment
are the most important characteristics, because all other parameters are
under square root. Obviously, $\approx $100\% enrichment is very desirable%
\footnote{%
Let us consider two detectors with different masses ($m_1$, $m_2$) and
enrichment ($\delta _1,$ $\delta _2$). Supposing that their other
characteristics ($\eta $, $t$, $R$, $Bg$) are the same and requiring the
equal sensitivities ($T_{1/2}^{^{\prime }}=T_{1/2}^{^{\prime \prime }}$), we
can obtain the relation between masses and enrichment of the detectors $%
m_1/m_2=(\delta _2/\delta _1)^2$, which speaks for itself.}.

One could also require $\approx $100\% detection efficiency, which is
possible, in fact, only for the ''active'' source technique. Indeed, the
strength of ''passive'' source can be enlarged by increasing its thickness,
which in turn lowers detection efficiency due to absorption of electrons in
the source, broadening and shifting of the $2\beta $ decay peak. To this
effect, the energy resolution of the detector is very essential because
events from the high energy tail of the continuous $2\nu $ distribution run
into the energy window of the $0\nu $ peak, generating background which
cannot be discriminated from the $0\nu $ signal\footnote{%
In both cases all features of events are similar: two electrons with the
same energies and identical angular distribution are emitted from one point
of the source simultaneously.}. Better energy resolution minimizes the $2\nu 
$ tail falling within the $0\nu $ interval, hence lowering this irreducible
background.

All mentioned statements are illustrated by fig. 1, where results of model
experiment to study $2\beta $ decay of $^{100}$Mo are presented. The
simulation were performed with the help of GEANT3.21 package \cite{GEANT}
and event generator DECAY4 \cite{Decay4}. The following assumptions were
accepted: mass of $^{100}$Mo source is one kg; measuring time is 5 years;
half-lives of $^{100}$Mo $2\beta $ decay are $T_{1/2}$(2$\nu )$ $=$ $10^{19}$
yr and $T_{1/2}$(0$\nu )$ = $10^{24}$ yr. The initial $2\beta $ decay
spectra (shown in fig. 1a and 1b for different vertical scales) were
obtained with $^{100}$Mo nuclei contained in the ideal (''active'' source)
detector with 100\% efficiency, zero background and the energy resolution of 
{\it FWHM}$=$10 keV. On the next step the $^{100}$Mo source was introduced
in the same detector but in the form of foil (''passive'' source technique).
The simulated spectra are depicted in fig. 1c (thickness of $^{100}$Mo foil
is 15 mg/cm$^2$) and fig. 1d (60 mg/cm$^2$). Then, the energy resolution of
the detector ($FWHM$) was taken into account and results are shown in fig.
1e ($FWHM$ = 4\% at 3 MeV) and fig. 1f ($FWHM$ = 8.8\% at 3 MeV). It is
evident from fig. 1 that ''passive'' source technique is not appropriate for
the observation of 0$\nu 2\beta $ decay with ratio of $T_{1/2}$(0$\nu )$ to $%
T_{1/2}$(2$\nu )$ more than $10^{5}$.

\nopagebreak
\begin{figure}[ht]
\begin{center}
\mbox{\epsfig{figure=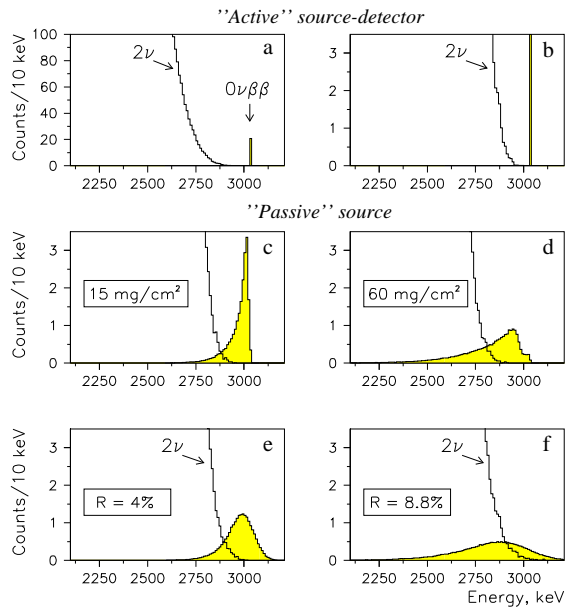,height=8.0cm}}
\caption {Simulated spectra of the model $2\beta $ decay experiment (5
yr measuring time) with 1 kg of $^{100}$Mo. (1a, 1b) ''Active'' source
technique: $^{100}$Mo nuclei in a detector with 100\% efficiency, zero
background, and with 10 keV energy resolution. (1c, 1d) ''Passive'' source
technique: $^{100}$Mo source in the same detector with foil thickness 15
mg/cm$^2$ (1c) and 60 mg/cm$^2$ (1d). (1e) The same as (1c) but the energy
resolution ($FWHM$) of the detector at 3 MeV is 4\%. (1f) The same as (1d)
but with $FWHM=8.8$\%.}
\end{center}
\end{figure}

Hence, we conclude that ''active'' source approach provides 4$\pi $ geometry
for the source, absence of self-absorption, and better energy resolution,
which does not depend on the angular and energy distribution of the
electrons emitted in $2\beta $ decay. These advantages of ''active''
detectors were understood a long ago and the first experiment of this type
was performed in 1966 by using $^{48}$CaF$_2 $ scintillator to study $2\beta 
$ decay of $^{48}$Ca \cite{Mat66}. In the next year semiconductor Ge(Li)
crystal was applied in the quest for $2\beta $ decay of $^{76}$Ge \cite
{Fio67}. Due to high purity and good energy resolution of the Ge(Li)
detectors the first valuable result with $^{76}$Ge ($T_{1/2}^{0\nu }\geq 1$0$%
^{21}$ yr) was obtained in 1970 \cite{Fio70}. After 30 years of strong
efforts this limit was advanced up to $T_{1/2}^{0\nu }\geq 1$0$^{25}$ yr in
the two current experiments performed by the IGEX \cite{IGEX} and
Heidelberg-Moscow \cite{Ge76} collaboration.

The IGEX is operating three 2-kg HP Ge detectors (enriched in $^{76}$Ge to $%
\approx $88\%) in the Canfranc Underground Laboratory (Spain). The shield
consists of 2.5 tons of archeological and 10 tons of 70-yr-old low-activity
lead, and plastic scintillator as cosmic muon veto. The pulse shape
discrimination techniques is applied to data. The background rate is equal
to $\approx $0.06 counts/yr$\cdot $kg$\cdot $keV (within the energy interval
2.0--2.5 MeV). The combined energy resolution for the $0\nu $2$\beta $ peak (%
$Q_{\beta \beta }=$ 2038.7 keV) is 4 keV. Analysis of 116.75 mole-years (or
8.87 kg$\times $yr in $^{76}$Ge) of data yields a lower bound $T_{1/2}^{0\nu
}\geq 1.5$7$\times $1$0^{25}$ yr at 90\% C.L. \cite{IGEX}.

The Heidelberg-Moscow experiment in the Gran Sasso Underground Laboratory
uses five HP Ge detectors (enriched in $^{76}$Ge to 86\%) with the total
active mass of 10.96 kg (125.5 moles of $^{76}$Ge). The passive and active
shielding, as well as pulse shape analysis (PSA) of data allows to reduce
background rate in the energy region of interest to the value of $\approx $%
0.06 counts/yr$\cdot $kg$\cdot $keV. The energy resolution at the energy of
2038.7 keV is 3.9 keV. After 24 kg$\times $yr of data with PSA a lower
half-life limit $T_{1/2}^{0\nu }\geq 1.$6$\times $1$0^{25}$ yr with 90\%
C.L. has been set for $^{76}$Ge \cite{Ge76}.

Therefore, on the basis of this brief analysis of the present status of 2$%
\beta $ decay experiments, we can formulate the following requirements to
the future ultimate sensitivity projects:

(i) The most sensitive $0\nu $ limits were reached with the help of
''active'' source method ($^{76}$Ge, $^{116}$Cd, $^{130}$Te, $^{136}$Xe),
thus one can suppose that future projects will belong to the same kind of
technique because only in this case the detection efficiency could be close
to 100\%.

(ii) The best $^{76}$Ge results were obtained by using $\approx $10 kg of
enriched detectors, hence, to reach the required level of sensitivity one
has to exploit the enriched sources with masses of hundreds kg. The latest
restricts the list of candidate nuclei because a large mass production of
enriched materials is possible only for several of them. These are $^{76}$%
Ge, $^{82}$Se, $^{116}$Cd, $^{130}$Te and $^{136}$Xe, which could be
produced by means of centrifugal separation\footnote{%
As it is known, the centrifugal isotope separation requires the substances
to be in gaseous form, thus xenon gas can be used directly. There also exist
volatile germanium, selenium, molybdenum and tellurium hexafluorides, as
well as metal--organic cadmium--dimethyl compound \cite{Isotop}.} and
therefore with a reasonable price \cite{Isotop}.

(iii) Because of the square root dependence of the sensitivity versus source
mass, it is not enough, however, to increase detector mass alone (even by
two orders of magnitude). The background should be also reduced down
substantially (practically to zero).

(iv) As it is obvious from fig. 1, the energy resolution is a crucial
characteristic, and for the challenging projects the $FWHM$ value cannot be
worse than $\approx $4\% at $Q_{\beta \beta }$ energy.

(v) It is anticipated that measuring time of the future experiments will be
of the order of $\approx $10 yr, hence detectors and set ups should be as
simple as possible to provide stable and reliable operation during such a
long period.

Evidently, it could be very difficult to find the project and to build up
the experiment, which would completely satisfy these severe requirements.
However, perhaps some of recent proposals could do it to a great extent,
thus let us consider them briefly.

An interesting approach to study 2$\beta $ decay of $^{136}$Xe ($Q_{\beta
\beta }=2468$ keV) makes use of the coincident detection of $^{136}$Ba$^{2+}$
ions (the final state of the $^{136}$Xe decay on the atomic level) and the 0$%
\nu $2$\beta $ signal with the energy of 2.5 MeV in a time projection
chamber (TPC) filled with liquid or gaseous Xe \cite{Moe91,Miya91,Miya96}.
Recently, the EXO project has been considered \cite{EXO}, where the
resonance ionization spectroscopy for the $^{136}$Ba$^{2+}$ ions
identification would be applied in a 40 m$^3$ TPC operated at 5--10 atm
pressure of enriched xenon ($\approx $1--2 tons of $^{136}$Xe). Estimated
sensitivity to neutrino mass is $\approx $0.01 eV \cite{EXO}. Another
proposal (originated from \cite{Ragh94}) is to dissolve $\approx $80 kg ($%
\approx $1.5 tons) of enriched (natural) Xe in the liquid scintillator of
the BOREXINO Counting Test Facility (CTF) in order to reach the $%
T_{1/2}^{0\nu }$ limit in the range of 10$^{24}$--10$^{25}$ yr \cite{Xe-CTF}.

The project MOON aims to make both the study of 0$\nu $2$\beta $ decay of $%
^{100}$Mo ($Q_{\beta \beta }=3034$ keV) and the real time studies of low
energy solar $\nu $ by inverse $\beta $ decay \cite{MOON}. The detector
module will be composed of $\approx $60,000 plastic scintillators (6 m$%
\times $0.2 m$\times $0.25 cm), the light outputs from which are collected
by 866,000 wave length shifter fibers ($\oslash $1.2 mm $\times $ 6 m),
viewed through clear fibers by 6800 16-anode photomultiplier tubes. The
proposal calls for the use of 34 tons of natural Mo (i.e. 3.3 tons of $%
^{100} $Mo) per module in the form of foil ($\approx $50 mg/cm$^2)$. The
sensitivity of such a module to the neutrino mass could be of the order of $%
\approx $ 0.05 eV \cite{MOON}.

The $^{160}$Gd ($Q_{\beta \beta }=1730$ keV) is an attractive candidate due
to large natural abundance (21.9\%), allowing to construct sensitive
apparatus with natural Gd$_2$SiO$_5$:Ce crystal scintillators (GSO). The
large scale experiment with $^{160}$Gd by using the GSO multi-crystal array
with the total mass of one-two tons ($\approx $200--400 kg of $^{160}$Gd) is
suggested with the projected sensitivity to the Majorana neutrino mass $%
\approx 0.04$ eV \cite{Gd-160}.

Using future large scale Yb-loaded liquid scintillation detectors for solar
neutrino spectroscopy \cite{Rag97} it is supposed to search for $2\beta ^{-}$
decay of $^{176}$Yb ($Q_{\beta \beta }=1087$ keV) and $\varepsilon \beta
^{+} $ decay of $^{168}$Yb ($Q_{\beta \beta }=1422$ keV). With about 20 tons
of natural Yb ($\approx $2.5 tons of $^{176}$Yb) the limit $T_{1/2}^{0\nu
}\geq $ 10$^{26}$ yr could be set on 0$\nu 2\beta $ decay of $^{176}$Yb ($%
m_\nu \leq $ 0.1 eV) \cite{Zub00}.

However, we recall that all mentioned projects require a significant amount
of R\&D to demonstrate their feasibility, thus the strong efforts and
perhaps long time will be needed before their realization. To this effect,
we offer the following safer proposals.

First of all, there are two projects NEMO-3 \cite{Pig99} and CUORICINO \cite
{Fio98} under construction now. The sensitivity of the NEMO-3 tracking
detector with a passive 10 kg of $^{100}$Mo source would be on the level of $%
\approx $4$\times $10$^{24}$ yr ($m_\nu $ $\leq $ 0.3--0.5 eV) \cite{Nemo00}.

The CUORICINO set up consists of 60 low temperature bolometers made of TeO$%
_2 $ crystals (750 g mass each) and is designed as a pilot step for a future
CUORE project for the $2\beta $ decay quest of $^{130}$Te with the help of
one thousand TeO$_2$ bolometers (total mass of 750 kg), which could reach $%
\approx $0.05 eV neutrino mass bound \cite{Fio98, Ger00}.

Recently a project CAMEO has been suggested \cite{CAMEO}, where the
super-low background and large sensitive volume of the already existing CTF
are used to study $^{116}$Cd. With $\approx $100 kg of enriched $^{116}$CdWO$%
_4$ crystal scintillators placed in the liquid scintillator of the CTF the
calculated sensitivity (in terms of the $T_{1/2}^{0\nu }$ limit) is $\approx 
$10$^{26}$ yr, which translates to the neutrino mass bound $m_\nu \leq 0.06$
eV. Similarly with one ton of $^{116}$CdWO$_4$ crystals located in the
BOREXINO apparatus (under construction) the constraint on the neutrino mass
can be pushed down to $m_\nu \leq 0.02$ eV \cite{CAMEO}.

Besides, two large scale projects for the 2$\beta $ decay quest of $^{76}$Ge
(MAJORANA \cite{MAJOR} and GENIUS \cite{GENIUS-98}) are proposed, which we
are going to discuss in more details.

{\bf MAJORANA.} The idea of this\ proposal is to use 210 HP Ge (enriched in $%
^{76}$Ge to $\approx $ 86\%) semiconductor detectors ($\approx $2.4 kg mass
of one crystal), which are placed in ''conventional'' super-low background
cryostat (21 crystals in one cryostat) \cite{MAJOR}. The detectors are
shielded by HP lead or copper. Each crystal will be supplied with six
azimuthal and two axial contacts, hence a proper spatial information will be
available for detected events. It is anticipated that segmentation of
crystals and pulse shape analysis (PSA) of data would reduce background rate
of the detectors to the level of $\approx $0.01 counts/yr$\cdot $kg$\cdot $%
keV at the energy 2 MeV, that is 6 times lower than that already reached in
the most sensitive $^{76}$Ge experiments \cite{Ge76,IGEX}. Thus, after 10 yr
of measurements $\approx $200 background counts will be recorded in the
vicinity of 0$\nu $2$\beta $ decay peak ($\approx $ 4 keV energy interval) 
\cite{MAJOR}. On this basis the half-life limit, $T_{1/2}$, can be
determined with the help of formula: ~lim~$T_{1/2}=\ln 2\cdot \eta \cdot
N\cdot t/\lim S$, where $N$ is the number of $^{76}$Ge nuclei ($N$ = 3.5$%
\times $10$^{27}$) and $\lim S$ is the maximal number of $0\nu 2\beta $
events which can be excluded with a given confidence level. To estimate
value of $\lim S$ we can use so called ''one (1.6; 2) $\sigma $ approach'',
in which the excluded number of effect's events is determined simply as
square root of the number of background counts in the energy region of
interest, multiplied by parameter (1; 1.6 or 2) in accordance with the
confidence level chosen (68\%, 90\% or 95\%). Notwithstanding its simplicity
this method gives the right scale of the sensitivity of any experiment.
Applying it to the projected MAJORANA data, one can get $\lim S$ $\approx $
20 counts at 90\% C.L., and whereby the bound $T_{1/2}$ $\approx $1$0^{27}$
yr. Depending on the nuclear matrix elements calculations used (see for
refs. \cite{Fae98,Suh98,Ge76}), it leads to the interval of the neutrino
mass limit $m_\nu \leq $ 0.05 -- 0.15 eV.

{\bf GENIUS.} The project intends to operate one ton of HP Ge (enriched in $%
^{76}$Ge to $\approx $ 86\%) semiconductor detectors \cite{GENIUS-98}. It is
scheduled that background of the GENIUS set up would be reduced by $\approx $%
200 times as compared with that of present experiments \cite{Ge76,IGEX}. To
reach this goal, ''naked'' Ge crystals will be placed in an extremely
high-purity liquid nitrogen (LN$_2$), which simultaneously serves as cooling
medium and shielding for the detectors.

The feasibility of operating Ge detectors in liquid nitrogen was
demonstrated by the measurements with three HP Ge crystals (mass of $\approx 
$0.3 kg each) \cite{Bau99}. With the 6 m cables between detectors (placed on
a common plastic holder inside liquid nitrogen) and outer preamplifiers the
energy threshold of $\approx $2 keV and the energy resolution of $\approx $1
keV (at 300 keV) were obtained \cite{Bau99}. The second question -- is it
indeed achievable to obtain so extremely low background level -- has been
answered by means of the Monte Carlo simulations. The latest were
independently performed by MPI, Heidelberg \cite{GENIUS-98} and INR, Kiev 
\cite{Pon98} groups. In accordance with simulations \cite{GENIUS-98,Pon98}
the necessary dimensions of the liquid nitrogen shield, which could fully
suppress the radioactivity from the surroundings (as that measured, for
instance, in the Gran Sasso Underground Laboratory) should be about 12 m in
diameter and 12 m in height. The required radioactive purity of the liquid
nitrogen should be at the level of $\approx $10$^{-15}$ g/g for $^{40}$K and 
$^{238}$U, $\approx $5$\times $1$0^{-15}$ g/g for $^{232}$Th, and 0.05 mBq/m$%
^3$ for $^{222}$Rn \cite{GENIUS-98,Pon98}. All these requirements (except
for radon) are less stringent than those already achieved in the BOREXINO
CTF: (2--5)$\times $1$0^{-16}$ g/g for $^{232}$Th and $^{238}$U
contamination in the liquid scintillators \cite{Bel96}. Therefore
purification of the liquid nitrogen to satisfy the GENIUS demands seems to
be quite realistic. The only problem is the radon contamination, those
required value is about 20 times less than that measured in liquid nitrogen
as $\approx $1 mBq/m$^3$ \cite{Bel96}. The final conclusions are derived
that in the GENIUS experiment the total background rate of $\approx $ 0.2
counts/yr$\cdot $keV$\cdot $t could be obtained in the energy region of the $%
\beta \beta $ decay of $^{76}$Ge \cite{GENIUS-98,Pon98}. On this basis the
projected $T_{1/2}$ limit can be estimated similarly as for the MAJORANA
proposal. For 10 yr measuring time the value of $\lim S$ is equal $\approx $
5 counts (90\% C.L.), thus with 7$\times $10$^{27}$ nuclei of $^{76}$Ge the
bound $T_{1/2}$ $\approx $1$0^{28}$ yr could be achieved, which translates
to the neutrino mass constraints $m_\nu \leq $ 0.015 -- 0.05 eV.

However, to reach the scheduled sensitivity the GENIUS apparatus must
satisfy very stringent and in some cases contradicting demands. For example,
super-low background rate of detectors requires the ultra-high purity of
liquid nitrogen and large dimensions of the vessel ($\oslash 12\times 12$ m)
with the total mass of a LN$_2$ of $\approx $1000 t. Ultra-high purity of
liquid nitrogen means that continuous purification of LN$_2$ during a whole
running experiment is needed. The power of the LN$_2$ purification system
(and maintenance costs) strongly depends on the liquid nitrogen consumption,
which in turn depends on the quality of the LN$_2$ tank thermoinsulation.
The method of passive thermoinsulation with the help of the polyethylene
foam isolation of 1.2 m thick was accepted for the GENIUS set up \cite
{GENIUS-98}. Despite its simplicity, the disadvantage of this solution is a
large LN$_2$ consumption because of huge dimensions of the LN$_2$ tank (heat
losses through the walls are straight proportional to their square). First,
it leads to the substantial maintenance cost of the experiment. Secondly,
and more important, this solution makes it very difficult to keep the
required ultra-high purity of LN$_2$ during the whole running period. It is
because that evaporation of LN$_2$ is the method of purification, thus pure
vapor will leave vessel, while all impurities will be kept in remaining LN$%
_2 $. In case of a large liquid nitrogen consumption this process will lead
to a permanent increasing of the LN$_2$ contamination level. Therefore, it
is clear that production, purification, operation and maintenance (together
with safety requirements) of more than one kiloton of ultra-high purity
liquid nitrogen in underground laboratory would require additional efforts
and lead to considerable costs and time for realization of the GENIUS
project.

With the aim to overcome all mentioned difficulties and make realization of
the high sensitivity $^{76}$Ge experiment simpler, the GEM project is
presented below.

\section{The GEM design and background simulation}

The GEM design is based on the following keystone ideas:

(a) ''Naked'' HP Ge detectors (enriched in $^{76}$Ge to 86 -- 90\%) are
operating in the ultra-high purity liquid nitrogen serving as cooling medium
and the first layer of shield simultaneously.

(b) Liquid nitrogen is contained in the vacuum cryostat made of HP copper.
The dimensions of the cryostat and consequently the volume of liquid
nitrogen are as minimal as necessary to eliminate the contribution of the
radioactive contaminations of the Cu cryostat to the background of HP Ge
detectors.

(c) The shield is composed of two parts: (i) inner shielding -- ultra-high
purity liquid nitrogen, whose contaminations are less than $\approx $10$%
^{-15}$ g/g for $^{40}$K and $^{238}$U, $\approx $5$\times $1$0^{-15}$ g/g
for $^{232}$Th, and 0.05 mBq/m$^3$ for $^{222}$Rn; (ii) outer part -- high
purity water, whose volume is large enough to suppress any external
background to the negligible level.

The optimization of the set up design as well as background simulation for
the GEM experiment were performed with the help of GEANT3.21 package and
event generator DECAY4. Scheme of the GEM device created on the basis of
simulation is shown in fig. 2. About 400 enriched HP Ge detectors ($\oslash $%
8.5$\times $8.5 cm, weight of $\approx $2.5 kg each) are located in the
center of a copper sphere (inner enclosure of the cryostat) with diameter
4.5 m and 0.6 cm thick, which is filled with liquid nitrogen. The detectors,
arranged in nine layers, occupied space of $\approx $90 cm in diameter. It
is supposed that crystals are fixed with the help of holder-system made of
nylon strings. The thin copper wire $\oslash 0.2$ mm is attached to each
detector to provide signal connection.

\nopagebreak
\begin{figure}[ht]
\begin{center}
\mbox{\epsfig{figure=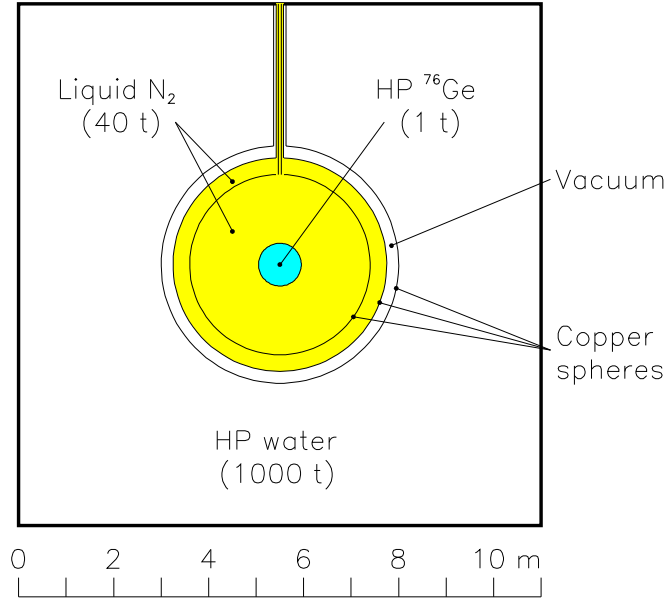,height=8.0cm}}
\caption {The scheme of the GEM set up.}
\end{center}
\end{figure}

The outer encapsulation of the cryostat with diameter 5 m is also made of HP
Cu with 0.6 cm thickness. Both enclosures of cryostat are connected by two
concentric copper pipes with outer vacuum pump, which maintains $\approx $10$%
^{-6}$ torr pressure in the space between two walls of the cryostat. The
latest (in combination with several layers of $\approx $5 $\mu $m thick
aluminized mylar film enveloping the inner Cu vessel and serving as thermal
radiation reflector) allows to reduce heat current through the walls of the
cryostat to the value of $\approx $2.5 W/m$^2$ \cite{Vacuum}, thus total
heat losses (including heat conduction through pipes, support structure and
cables) are near 200 W. This corresponds to a reasonable LN$_2$ consumption
of about 150 kg per day.

Moreover, to provide the most stable and quiet operation of HP Ge detectors,
the volume with liquid nitrogen is divided in turn into two zones with the
help of an additional Cu sphere with diameter 3.8 m and 1 mm thick. The HP
Ge detectors are contained in the latest, where only tiny fraction of heat
current through thin signal cables and holder strings could reach this
volume. Outer LN$_2$ zone between inner wall of the cryostat and sphere with
Ge crystals would serve as additional and very efficient thermal shield \cite
{Vacuum}. Hence, LN$_2$ consumption in the inner volume with detectors would
be extremely low, which allows to keep there the ultra-high purity of LN$_2$
and stable operation conditions for a whole running period.

Another important advantage of the proposed solution is that detectors are
located inside a module, and all procedures of cleaning, mounting of
crystals, testing, etc. can be performed in special clean room with all
available precautions to avoid any contaminations of the detectors and the
inner vessel.

The cryostat is placed into the HP ($\approx $10$^{-14}$ g/g for $^{40}$K, $%
^{232}$Th, $^{238}$U, and $\approx $10 mBq/m$^3$ for $^{222}$Rn) water
shield with mass $\approx $1000 t contained in the steel tank $\oslash $11$%
\times $11 m. We remind that even slightly better radio-purity levels were
already achieved for the water shield of the BOREXINO CTF operating in the
Gran Sasso Underground Laboratory \cite{Bel96}. The dimensions of the CTF
water tank are practically the same ($\oslash $11$\times $10 m), hence this
shield could be used for the GEM experiment easily. Because water is a
Cherenkov medium with excellent optical properties, such a shield equipped
with limited number of photomultipliers would serve as additional veto
system for muons in the GEM detector.

The developed design of the GEM set up reduces the dimensions of the LN$_2$
volume substantially and allows to solve the problems of thermoinsulation,
ultra-high purity conditions, LN$_2$ consumption, safety requirements, etc.

\subsection{Background simulations}

In the calculations the model of the GEM experiment described above was used
(see fig. 2). The total mass of detectors is equal $\approx $1 t, liquid
nitrogen -- $\approx $40 t, copper cryostat -- $\approx $7 t, water shield
-- 1000 t, holder-system -- $\approx $2 kg, and copper wires -- $\approx $1
kg. As it is already mentioned the simulation of the background and in
particular the decay of various radioactive nuclides in the installation was
performed with the help of GEANT3.21 package and event generator DECAY4. The
energy threshold of the HP Ge detectors was set to 1 keV and only single
signals in one of all detectors (anti-coincidence mode) were taken into
account. The origins of background can be divided into internal and external
ones. Internal background arises from residual impurities in the crystal
holder system, in the Ge crystals themselves, in the liquid nitrogen, copper
cryostat, water, in the steel vessel, and from activation of all mentioned
materials at the Earth surface. External background is generated by events
originating outside the shield, such as photons and neutrons from the Gran
Sasso rock, muon interactions and muon induced activities.

\subsubsection{Radioactive impurities of the detectors and materials{\bf \ }}

The values of radioactive contamination of the Ge detectors and materials
used (liquid nitrogen, copper wires and cryostat, water, steel vessel) by $%
^{40}$K and nuclides from natural radioactive chains of $^{232}$Th and $%
^{238}$U, accepted for our calculations, are listed in Table 1. Possible
contamination of $^{76}$Ge crystals were calculated by using the data of the
Heidelberg-Moscow experiment with $^{76}$Ge detectors \cite{Ge76,Gun97}. The
absence of any $\alpha $ peaks in the measured spectra (for 17.7 kg$\times $%
yr statistics) leads to the upper limits (90\% C.L.) presented in Table 1.
Data on purity of copper for $^{40}$K, $^{232}$Th and $^{238}$U are taken
from \cite{Gun97}. The copper cosmogenic activities of $^{54}$Mn (23 $\mu $%
Bq/kg), $^{57}$Co (30 $\mu $Bq/kg), $^{58}$Co (50 $\mu $Bq/kg), $^{60}$Co
(70 $\mu $Bq/kg), as well as anthropogenic activities of $^{125}$Sb (50 $\mu 
$Bq/kg), $^{207}$Bi (8 $\mu $Bq/kg), $^{134}$Cs (150 $\mu $Bq/kg) and $%
^{137} $Cs (11 $\mu $Bq/kg) are accepted on the basis of measurements with
the Ge detectors of the Heidelberg-Moscow experiment \cite{Gun97}. For
steel, the upper limits from \cite{Jag93} are assumed. For the water the
actual radio-purity levels obtained in the already operated BOREXINO water
plant \cite{Bel96} are quoted in Table 1. The radiopurity criteria supposed
for the liquid nitrogen ($\approx $10$^{-15}$ g/g for $^{40}$K and $^{238}$%
U, $\approx $5$\times $1$0^{-15}$ g/g for $^{232}$Th) seem to be realistic
in light of the results already achieved by the BOREXINO collaboration for
the purity of the liquid scintillators: (2--5)$\times $1$0^{-16}$ g/g for $%
^{232} $Th and $^{238}$U \cite{Bel96}. Moreover, due to recent development
of the liquid nitrogen purification system for the BOREXINO experiment \cite
{Heus00} the $^{222}$Rn contamination of the liquid nitrogen was also
reduced down to the level of $\approx $1 $\mu $Bq/m$^3$ \cite{Heus00}, which
is lower than our requirement $\approx $50 $\mu $Bq/m$^3$. For the
radiopurity of the holder-system we assume the value of 10$^{-12}$ g/g for
the U/Th decay chains, which is already achieved by the SNO collaboration
for the acrylic \cite{SNO}.

\begin{table}[tbp]
\caption{Radioactive impurities of the detectors and materials accepted for
the simulation}
\begin{center}
\begin{tabular}{|l|l|l|l|l|}
\hline
\multicolumn{1}{|l|}{Materials} & $^{40}$K, & $^{232}$Th, & 
\multicolumn{1}{|l|}{$^{238}$U,} & $^{222}$Rn, \\ 
\multicolumn{1}{|l|}{(mass)} & g/g & g/g & \multicolumn{1}{|l|}{g/g} & mBq/m$%
^3$ \\ \hline
\multicolumn{1}{|l|}{HP $^{76}$Ge (1 t)} & -- & 5.7$\times $10$^{-15}$ & 
\multicolumn{1}{|l|}{1.8$\times $10$^{-15}$} & -- \\ 
\multicolumn{1}{|l|}{Liquid N$_2$ (40 t)} & 1.0$\times $10$^{-15}$ & 5.0$%
\times $10$^{-15}$ & \multicolumn{1}{|l|}{1.0$\times $10$^{-15}$} & 0.05 \\ 
\multicolumn{1}{|l|}{Holder system (2 kg)} & -- & 1.0$\times $10$^{-12}$ & 
\multicolumn{1}{|l|}{1.0$\times $10$^{-12}$} & -- \\ 
\multicolumn{1}{|l|}{Copper wires (1 kg)} & 4.5$\times $10$^{-10}$ & 3.0$%
\times $10$^{-12}$ & \multicolumn{1}{|l|}{5.4$\times $10$^{-12}$} & -- \\ 
\multicolumn{1}{|l|}{and vessels (7 t)} &  &  & \multicolumn{1}{|l|}{} &  \\ 
\multicolumn{1}{|l|}{Water (1000 t)} & 1.0$\times $10$^{-14}$ & 1.0$\times $%
10$^{-14}$ & \multicolumn{1}{|l|}{1.0$\times $10$^{-14}$} & 10 \\ 
\multicolumn{1}{|l|}{Steel vessel (90 t)} & 5.0$\times $10$^{-10}$ & 1.0$%
\times $10$^{-9}$ & \multicolumn{1}{|l|}{1.0$\times $10$^{-9}$} & -- \\ 
\hline
\end{tabular}
\end{center}
\end{table}

The full decay chains were simulated in assumption of the chains
equilibrium. Results of calculation are presented in Table 2. For internal
impurities in HP Ge detectors, two values are given: without (in square
brackets) and with time-amplitude analysis of events, where information
about the energies and arrival time of each event is used for analysis and
selection of some decay chains in U and Th families (see, f.e., ref. \cite
{Cd2000}).

\begin{table}[tbp]
\caption{Calculated background rate of the detectors at the energy 2038 keV
due to internal impurities of the materials. For the LN$_2$ and water the $%
^{222}$Rn contributions are included in column for $^{238}$U}
\begin{center}
\begin{tabular}{|llll|}
\hline
\multicolumn{1}{|l|}{Material} & Background & rate at 2 MeV, & counts/yr$%
\cdot $keV$\cdot $t \\ \cline{2-4}
\multicolumn{1}{|l|}{} & \multicolumn{1}{l|}{$^{232}$Th} & 
\multicolumn{1}{l|}{$^{238}$U} & total \\ \hline
\multicolumn{1}{|l|}{HP $^{76}$Ge} & \multicolumn{1}{l|}{2.0$\times $10$%
^{-3} $} & \multicolumn{1}{l|}{4.3$\times $10$^{-3}$} & 6.3$\times $10$^{-3}$
\\ 
\multicolumn{1}{|l|}{} & \multicolumn{1}{l|}{[4.6$\times $10$^{-2}$]} & 
\multicolumn{1}{l|}{[1.6$\times $10$^{-1}$]} & [2.1$\times $10$^{-1}$] \\ 
\multicolumn{1}{|l|}{Liquid N$_2$} & \multicolumn{1}{l|}{5.3$\times $10$%
^{-3} $} & \multicolumn{1}{l|}{1.2$\times $10$^{-2}$} & 1.7$\times $10$^{-2}$
\\ 
\multicolumn{1}{|l|}{Holder system} & \multicolumn{1}{l|}{2.0$\times $10$%
^{-3}$} & \multicolumn{1}{l|}{1.4$\times $10$^{-2}$} & 1.6$\times $10$^{-2}$
\\ 
\multicolumn{1}{|l|}{Cu wires} & \multicolumn{1}{l|}{1.5$\times $10$^{-4}$}
& \multicolumn{1}{l|}{2.5$\times $10$^{-3}$} & 2.7$\times $10$^{-3}$ \\ 
\multicolumn{1}{|l|}{Inner Cu sphere $\oslash $3.8m} & \multicolumn{1}{l|}{
4.3$\times $10$^{-3}$} & \multicolumn{1}{l|}{3.0$\times $10$^{-3}$} & 7.3$%
\times $10$^{-3}$ \\ 
\multicolumn{1}{|l|}{Two Cu cryostat walls} & \multicolumn{1}{l|}{1.9$\times 
$10$^{-2}$} & \multicolumn{1}{l|}{8.6$\times $10$^{-3}$} & 
\multicolumn{1}{l|}{2.8$\times $10$^{-2}$} \\ 
\multicolumn{1}{|l|}{Water} & \multicolumn{1}{l|}{3.0$\times $10$^{-3}$} & 
\multicolumn{1}{l|}{2.0$\times $10$^{-3}$} & 5.0$\times $10$^{-3}$ \\ 
\multicolumn{1}{|l|}{Steel vessel} & \multicolumn{1}{l|}{1.4$\times $10$%
^{-3} $} & \multicolumn{1}{l|}{--} & 1.4$\times $10$^{-3}$ \\ \hline
\multicolumn{1}{|l|}{Total} & \multicolumn{1}{l|}{3.7$\times $10$^{-2}$} & 
\multicolumn{1}{l|}{4.7$\times $10$^{-2}$} & {8.4$\times $10$^{-2}$ } \\ 
\hline
\end{tabular}
\end{center}
\end{table}

It is obvious from Table 2 that two Cu enclosures of the cryostat and holder
system give the main contribution to background. Besides, the results of
simulations show that demands to the purity of water shield can be lowered
to the level of about $10^{-13}$ g/g for U (Th) contaminations. The latest
means that maintenance costs of the GEM experiment can be lowered too.

\subsubsection{Cosmogenic activities in HP $^{76}$Ge detectors}

To estimate the cosmogenic activity produced in the HP Ge crystals, the
program COSMO \cite{Mar92} was used. This code calculates the production of
all radionuclides with half-lives in the range of 25 days -- 5 million years
by nucleon-induced reactions in a given target, taking into account the
variation of spallation, evaporation, fission and peripheral reaction
cross-sections with nucleon energy, target and product charge and mass
numbers, as well as energy spectrum of cosmic ray nucleons near the Earth's
surface \cite{Mar92}.

Cosmogenic activities in Ge were calculated for HP Ge detectors enriched in $%
^{76}$Ge to 86\% (other Ge isotopes: $^{70}$Ge -- 3.2\%, $^{72}$Ge -- 4.2\%, 
$^{73}$Ge -- 1.2\%, $^{74}$Ge -- 5.4\%). An activation time of 30 days at
sea level\footnote{%
We supposed that Ge materials and crystals were additionally shielded
against activation during production and transportation. For example, 20 cm
of Pb would lower the cosmic nucleons flux by one order of magnitude that
means the same reduction factor for the most of cosmogenic activities.}, and
a deactivation time of three years in underground were assumed. From the
total number of 41 nuclides with $T_{1/2}\geq 25$ d produced in Ge crystals,
we present in Table 3 the most dangerous ones which give the noticeable
background near the energy 2038 keV ($Q_{\beta \beta }$ value of $^{76}$Ge).
For $^{68}$Ga activity two values are given: without (in square brackets)
and with time-amplitude analysis of events.

\begin{table}[tbp]
\caption{Cosmogenic activities produced in HP $^{76}$Ge detectors. The
background rate at the energy 2038 keV is averaged during one year period of
data taking}
\begin{center}
\begin{tabular}{|l|l|l|l|}
\hline
Nuclide & Mode of decay & Activity & Background \\ 
&  & after 3 yr, & at 2 MeV, \\ 
($T_{1/2}$) & ($Q $, keV) & $\mu $Bq/kg & counts/yr$\cdot $keV$\cdot $t \\ 
\hline
$^{22}$Na (2.6 yr) & EC/$\beta ^{+}$ (2842) & 2.0$\times $10$^{-3}$ & 3.5$%
\times $10$^{-3}$ \\ 
$^{46}$Sc (83.8 d) & $\beta ^{-}$\qquad (2367) & 2.5$\times $10$^{-5}$ & 1.0$%
\times $10$^{-4}$ \\ 
$^{56}$Co (78.8 d) & EC/$\beta ^{+}$ (4568) & 1.9$\times $10$^{-5}$ & 3.2$%
\times $10$^{-6}$ \\ 
$^{58}$Co (70.8 d) & EC/$\beta ^{+}$ (2308) & 6.0$\times $10$^{-5}$ & 1.4$%
\times $10$^{-5}$ \\ 
$^{60}$Co (5.27 yr) & $\beta ^{-}$\qquad (2824) & 6.6$\times $10$^{-2}$ & 5.4%
$\times $10$^{-2}$ \\ 
$^{68}$Ga (68.1 m) & EC/$\beta ^{+}$ (2921) & 5.0$\times $10$^{-2}$ & 0.018
[0.15] \\ \hline
Total &  &  & 0.07 [0.22] \\ \hline
\end{tabular}
\end{center}
\end{table}

It is clear from Table 3 that background at 2038 keV is caused mainly by $%
^{22}$Na, $^{60}$Co and $^{68}$Ga (a daughter of cosmogenic $^{68}$Ge).
Remaining $^{68}$Ga contribution could be suppressed significantly by using
the time-amplitude analysis due to specific features of the $^{68}$Ge $%
\rightarrow $ $^{68}$Ga decay chain. Indeed, 88\% of the electron captures
in $^{68}$Ge to the ground state of $^{68}$Ga result in the sharp 10.4 keV
peak (K capture). Using these events as triggers for time-amplitude analysis
of the subsequent counts during few half-lives of $^{68}$Ga ($T_{1/2}$ =
68.1 m), it is possible to remove up to 88\% of the remaining activity of $%
^{68}$Ga. The expected rate of $^{68}$Ge decay (one event per 60 days per
detector) would allow to use such an approach. The background from $^{60}$Co
can be also decreased by additional annealing of Ge crystals in the
underground laboratory. Preliminary study shows that $^{60}$Co can be
removed out of detectors due to its large diffusion mobility in Ge at high
temperatures \cite{Gri91}. All mentioned approaches will reduce the
cosmogenic background rate to the value less than 3$\times $1$0^{-2}$
counts/yr$\cdot $keV$\cdot $t near 2038 keV.

\subsubsection{External background}

There are several origins of external background for the proposed GEM
detector. These are neutrons and $\gamma $ quanta from natural environmental
radioactivity, cosmic muons ($\mu $ showers and muon induced neutrons,
inelastic scattering and capture of muons), etc. From all of them only $%
\gamma $ quanta from environment were simulated in the present work, while
others were simply estimated as negligible on the basis of the results of
ref. \cite{GENIUS-98,Pon98}, where such origins and contributions were
investigated carefully.

We simulated the influence of the photon flux with the energies up to 3 MeV
measured in hall C of the Gran Sasso laboratory \cite{arpesella92}, where
the main contributions are originating from U and Th contamination of
concrete walls. Among them mainly $\gamma $'s with the energy of 2614 keV
(flux $\approx $5$\times $1$0^9$ m$^{-2}\cdot $ yr$^{-1}$) can be dangerous
for the experiment. In our calculations approximately 10$^{15}$ external $%
\gamma $'s with E$_\gamma $=2614 keV were simulated, yielding the detector
background at the energy 2038 keV of about 0.01 counts/yr$\cdot $keV$\cdot $%
t. \hspace{1.0in}

Summarizing all background origins (internal and external) we get the total
background rate of the GEM experiment less than 0.2 counts/yr$\cdot $keV$%
\cdot $t (at 2038 keV). The simulated response functions of the GEM set up
after 10 yr measuring time for $2\beta $ decay of $^{76}$Ge ($T_{1/2}^{2\nu
} $ $=$1.8$\times $10$^{21}$ yr \cite{Gun97} and $T_{1/2}^{0\nu }$ $=$10$%
^{27}$ yr), as well as background contribution from contaminations of the
holder system and copper cryostat walls are depicted in fig. 3. It is
obvious from this figure that measured background at the energies below 1950
keV is dominated by two neutrino $2\beta $ decay distribution of $^{76}$Ge
(the total number of $\approx $2.6$\times $10$^7$ counts are recorded),
while at 2040 keV the main sources of background are contamination of the
holder system and copper cryostat walls by nuclides from U and Th chains. On
the other hand, it is also evident from fig. 3 that 0$\nu 2\beta $ decay of $%
^{76}$Ge with half-life of 10$^{27}$ yr would be clearly registered (there
are 42 counts in the $0\nu 2\beta $ decay peak).

\nopagebreak
\begin{figure}[ht]
\begin{center}
\mbox{\epsfig{figure=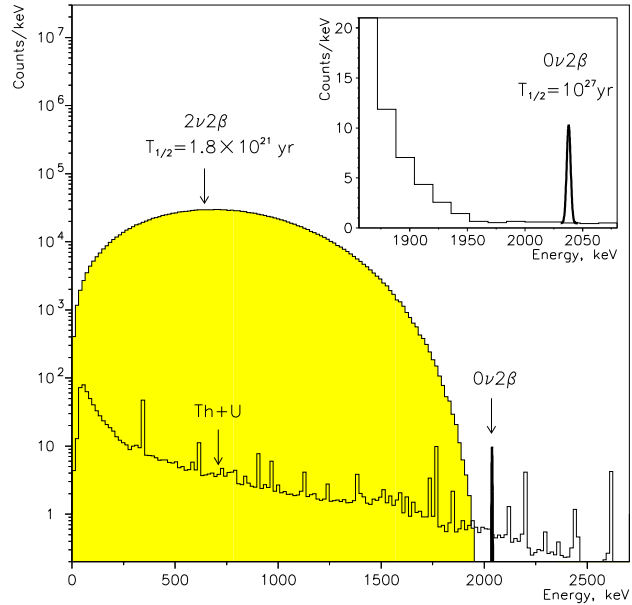,height=8.0cm}}
\caption {The response functions of the GEM-II set up with 1000 kg of HP 
$^{76}$Ge crystals and after 10 yr of measurements for $2\beta $ decay of $%
^{76}$Ge with $T_{1/2}^{2\nu }$ $=$1.8$\times $10$^{21}$ yr \cite{Gun97} and 
$T_{1/2}^{0\nu }$ $=$10$^{27}$ yr (solid histogram), as well as background
contribution from contaminations of the holder system and copper cryostat
walls by nuclides from $^{232}$Th and $^{238}$U families. In the insert the
summed spectrum in the vicinity of the 0$\nu 2\beta $ decay peak of $^{76}$%
Ge is shown in the linear scale.}
\end{center}
\end{figure}

The sensitivity of the GEM can be expressed in the same manner as for the
MAJORANA and GENIUS proposals (see Section 2). For 10 yr measuring period
the value of $\lim S$ is equal $\approx $5 counts (90\% C.L.), thus taking
into account the number of $^{76}$Ge nuclei (7$\times $10$^{27}$) and
detection efficiency ($\eta $ $\approx $ 0.95), the half-life bound $T_{1/2}$
$\approx $1$0^{28}$ yr could be achieved. Depending on the nuclear matrix
elements calculations \cite{Fae98,Suh98,Ge76}, the projected limit
corresponds to the following range of the neutrino mass constraints: $m_\nu
\leq $ 0.015 -- 0.05 eV.

The realization of the GEM experiment seems to be reasonably simple due to
fact that developed design of the set up has practically no technical risk.
To this effect, the very attractive feature of the project is the
possibility to use already existing BOREXINO CTF \cite{Bel96} as outer
shield, because the CTF fits all the GEM requirements concerning radiopurity
and dimensions of the water shield. In addition, one of forthcoming large
underground neutrino detectors such as KamLand \cite{KamLand} or BOREXINO
could be also appropriate for this purpose.

The cost of the GEM experiment is estimated as about 150 M\$, whose main
part would be for the production of enriched materials. However, we consider
that at the first phase of the project the measurements will be performed
with one ton of natural HP Ge detectors, whose cost (together with cost of
the cryostat) does not exceed 5 M\$. Beside of important technical tasks
which must be solved on the first stage of the GEM to prove feasibility of
the project and to test the developed design, the GEM-I phase with its
relatively modest cost would bring the outstanding physical results. Indeed,
in accordance with the formula for sensitivity of any 0$\nu 2\beta $ decay
experiment (see Sect. 2) the reachable half-life limit depends straight
proportional on the abundance or enrichment of candidate nuclei contained in
the detector. For the GEM-I the natural abundance of $^{76}$Ge (7.6\%) is
about 11 times smaller as compared with the enrichment supposed for the
second stage (86\%). Because any other characteristics of the set up ($\eta $%
, $m$, $t$, $R$, $Bg$) are the same for both GEM-I and GEM-II phases, the
half-life bound, which would be obtained with natural HP Ge detectors is
about one order of magnitude lower: $T_{1/2}$ $\approx $1$0^{27}$ yr. The
latest translates to the neutrino mass constrains $m_\nu \leq $ 0.05 eV,
which is also of great interest for many theoretical models.

Another and very important issue of the GEM-I stage is the quest for the
dark matter particles (see reviews \cite{Jun96,Ram99,Bau00a}). It has been
shown by Monte Carlo simulations \cite{GENIUS-98,Pon98} that for the GENIUS
project exploiting $\approx $100 kg of natural HP Ge detectors the
background rate of $\approx $40 counts/yr$\cdot $keV$\cdot $t could be
obtained in the low energy region (10--100 keV) relevant for the WIMP dark
matter study. The main contributions to this rate are from: (a) 2$\nu 2\beta 
$ decay of $^{76}$Ge with $T_{1/2}^{2\nu }$ $=$1.8$\times $10$^{21}$ yr \cite
{Gun97} ($\approx $20 counts/yr$\cdot $keV$\cdot $t ); (b) cosmogenic
activities in HP Ge crystals ($\approx $10 counts/yr$\cdot $keV$\cdot $t );
(c) internal radioactive contamination of the liquid nitrogen, copper wires
and holder system ($\approx $10 counts/yr$\cdot $keV$\cdot $t ). We
estimated that even lower background rate could be reached in the GEM-I set
up, where only inner volume with $\approx $200 kg of HP Ge detectors will be
used for the dark matter search, while outer layers with remaining $\approx $%
800 kg of HP Ge crystals would serve as super-high purity passive and active
shields for the inner detectors. Our simulation shows that in such a
configuration additional suppression of the background component (c) could
be obtained, which would allow to reach the highest sensitivity for the dark
matter search as compared with other projects (see f. e. refs. \cite
{Kla01,Klap01}).

\section{Implications of the high sensitivity 2$\beta $ decay experiments
and conclusions}

In this section we will discuss briefly the physical implications of future 2%
$\beta $ decay experiments, those sensitivity to neutrino mass limit would
be of the order of 0.05 eV (CAMEO, CUORE, EXO, GEM-I, MAJORANA, MOON, etc.)
and $\approx $0.01 eV (GEM-II, GENIUS).

As it was already mentioned in Introduction, many extensions of the Standard
Model incorporate lepton number violating interactions and thus could lead
to 0$\nu $2$\beta $ decay. Besides conventional left-handed neutrino
exchange mechanism of the 0$\nu $2$\beta $ decay, such theories offer many
other possibilities to trigger this process \cite{Fae98,Klap98,Suh98}.

For instance, in left-right symmetric GUT models neutrinoless 2$\beta $
decay can be mediated by heavy right-handed neutrinos \cite{Doi83}. It was
shown \cite{Kla97c} that 2$\beta $ decay experiments with the sensitivity $%
m_\nu \leq 0.01$ eV would be at the same time sensitive to right-handed $W_R$
boson masses up to $m_{W_R}\geq 8$ TeV (for a heavy right-handed neutrino
mass $\langle m_N\rangle =1$ TeV) or $m_{W_R}\geq 5.3$ TeV (for $\langle
m_N\rangle =m_{W_R}$). These limits, which therefore could be established
the by GEM-II experiment, are compared with those expected for LHC \cite
{Riz96}.

Another new type of gauge bosons predicted by some GUTs are leptoquarks
(LQ), which can transform quarks to leptons. Direct searches for leptoquarks
in deep inelastic $ep$-scattering at HERA give lower limits on their masses $%
M_{LQ}\geq 225-275$ GeV (depending on the LQ type and coupling) \cite{H196}.
Leptoquarks can induce 0$\nu $2$\beta $ decay via LQ-Higgs couplings, thus
restrictions on leptoquark masses and coupling constants can be derived \cite
{Hir96}. Detailed study performed in ref. \cite{MPI99} yields the conclusion
that GENIUS-like experiment would be able to reduce limit on LQ-Higgs
couplings down to $\approx 10^{-7}$ for leptoquarks with masses in the range
of 200 GeV. If no effect (0$\nu $2$\beta $ decay) will be found, it means
that either LQ-Higgs coupling must be smaller than $\approx 10^{-7}$ or
there exist no leptoquarks (coupling with electromagnetic strength) with
masses below $\approx 10$ TeV \cite{MPI99}.

Hypothetical substructure of quarks and leptons (compositeness) can also
give rise to a new 0$\nu $2$\beta $ decay mechanism by exchange of composite
heavy Majorana neutrinos \cite{Pan97}. Recent analysis \cite{Pan99} shows
that the most sensitive at present 0$\nu $2$\beta $ results with $^{76}$Ge 
\cite{Ge76,IGEX} yield the bound on the excited Majorana neutrino mass $%
m_N\geq 272$ GeV -- which already exceeds the ability of LEP-II to test
compositeness -- while future $^{76}$Ge experiments (GEM-II, GENIUS) would
shift this limit to $m_N\geq 1$ TeV competitive with the sensitivity of LHC 
\cite{Pan99}.

Moreover, there are also possible 0$\nu $2$\beta $ decay mechanisms based on
the supersymmetric (SUSY) interactions: exchange of squarks, etc., within $R$%
-parity violating \cite{Moh86,Hir95,Fae97,Wod99} and exchange of sneutrinos,
etc. in $R$-parity conserving SUSY models \cite{Hir97}. It allows 2$\beta $
decay experiments to enter in the field of supersymmetry, where competitive
restrictions on the sneutrino masses, $R$-parity violating couplings and
other parameters could be obtained \cite{Hir98b,Bha99}.

Now we are going to consider the relations between 0$\nu $2$\beta $ decay
studies and neutrino oscillation searches to demonstrate the role which
future 2$\beta $ experiments can play in the reconstruction of the neutrino
mass spectrum. At present this topic is widely discussed in literature, thus
interested readers are referred to the latest publications \cite
{Klap00,Bil99,Kla01,Vis99,Cza99,Cza00,Czak00,Bil01,Klab01,Klac01}, while we
will summarize the most important results very shortly.

There exist several schemes for the neutrino masses and mixing offered by
theoretical models on the basis of observed oscillation data for the solar
and atmospheric neutrinos \cite{Klap00,Bil99,Bil01}. These are schemes with:
normal and inverse neutrino mass hierarchy; partial and complete mass
degeneracy, as well as scenario with 4 neutrinos, etc. For each of these
schemes several solutions exist: (SMA) small mixing angle
Mikheyev-Smirnov-Wolfenstein (MSW) solution; (LMA) large mixing angle MSW
solution; (LOW) low mass MSW solution; (VO) vacuum oscillation solution.
Careful analysis of these schemes and solutions performed in refs. \cite
{Klap00,Bil99,Bil01} lead to the following statements: (a) effective
neutrino mass, $\langle m_\nu \rangle $, which is allowed by oscillation
data and could be observed in 2$\beta $ decay, is different for different
schemes and solutions, hence 2$\beta $ decay data could substantially narrow
or restrict this wide choice of possible models; (b) the whole range of
allowed $\langle m_\nu \rangle $ values is 0.001--1 eV, where there are
three key scales of $\langle m_\nu \rangle $: 0.1 eV; 0.02 eV and 0.005 eV.
If future 2$\beta $ decay experiments will prove that $\langle m_\nu \rangle 
$ $\geq $ 0.1 eV, then all schemes would be excluded, except those with
neutrino mass degeneracy or with 4 neutrinos and inverse mass hierarchy \cite
{Klap00}. With the $\langle m_\nu \rangle $ bound of about 0.02 -- 0.05 eV
several other solutions will be excluded, while if neutrino mass limit is $%
\langle m_\nu \rangle $ $\leq $ 0.005 eV the survived schemes are those with
mass hierarchy or with partial degeneracy. The following citation from \cite
{Bil01} emphasizes importance of the future 2$\beta $ decay searches: ''The
observation of the 0$\nu $2$\beta $ decay with a rate corresponding to $%
\langle m_\nu \rangle $ $\approx $ 0.02 eV can provide unique information on
the neutrino mass spectrum and on the CP-violation in the lepton sector, and
if CP-invariance holds -- on the relative CP-parities of the massive
Majorana neutrinos.''

Hence, it is obvious that GEM experiment will bring crucial results for the
reconstruction of the neutrino mass spectrum and mixing not only at its
final GEM-II stage with enriched detectors ($\langle m_\nu\rangle $ $\approx 
$ 0.015 eV), but also at the first phase with natural HP Ge crystals ($%
\langle m_\nu\rangle $ $\approx $ 0.05 eV). This statement is true for any
other topics discussed above.

Furthermore, namely the GEM-I with the realistic energy threshold of 10 keV
and with anticipated background rate of $\approx $ 40 counts/yr$\cdot $keV$%
\cdot $t below 100 keV would provide the highest sensitivity for the WIMP
dark matter search. It is demonstrated by the exclusion plot of the
WIMP-nucleon elastic scattering cross section for the GEM-I, which is
depicted in fig. 4 together with the best current and other projected limits%
\footnote{%
The serious background problem for the dark matter quest with Ge detectors
is cosmogenic activity of $^3$H produced in Ge \cite{GENIUS-98,Pon98,Klac01}%
. For the GEM-I we estimated the total $^3$H activity as $\approx $5000
decays/yr$\cdot $t, which is in good agreement with the result of \cite
{Klac01} and contributes $\approx $10 counts/yr$\cdot $keV$\cdot $t to the
total background rate in the energy interval 10-100 keV.}. The theoretical
prediction for allowed spin-independent elastic WIMP-proton scattering cross
section calculated in ref. \cite{Ell00} in the framework of the constrained
minimal supersymmetric standard model (MSSM) is also shown there\footnote{%
Very similar predictions from theoretical considerations in the MSSM with
relaxed unification condition were derived in ref. \cite{Bed01}.}. It is
obvious from fig. 4 that GEM-I would test the MSSM prediction by covering
the larger part of the predicted SUSY parameter space. In that sense GEM
experiment could be competitive even with LHC in the SUSY quest \cite{Riz96}%
. At the same time with fiducial mass of HP Ge detectors of $\approx $200 kg
GEM-I would be able to test and identify unambiguously (within one year of
data taking \cite{Ceb01}) the seasonal modulation signature of the dark
matter signal from the DAMA experiment \cite{DAMA00} by using an alternative
detector technology.

\nopagebreak
\begin{figure}[ht]
\begin{center}
\mbox{\epsfig{figure=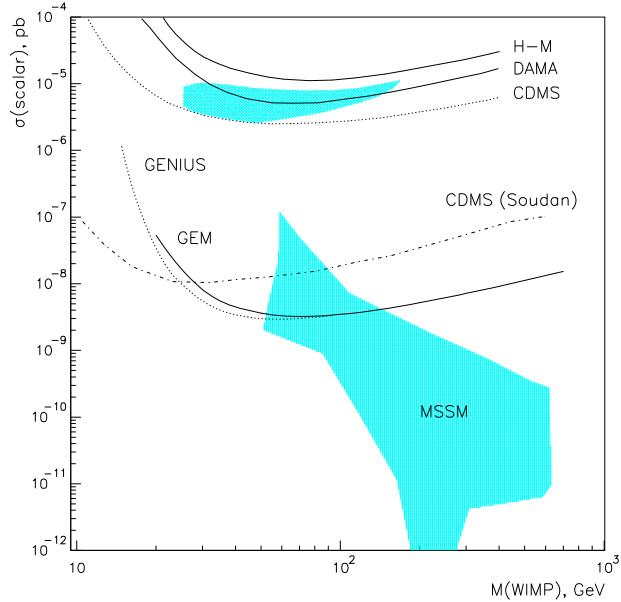,height=8.0cm}}
\caption {Exclusion plots of the spin-independent WIMP-nucleon elastic
cross section versus WIMP mass. The regions above the curves are excluded at
90\% C.L. Current limits from Heidelberg-Moscow (H-M) \cite{H-M99}, DAMA 
\cite{DAMA98} and CDMS \cite{CDMS00} experiments are shown in the upper part
of figure. The small shaded area: 2$\sigma$ evidence region from the DAMA
experiment \cite{DAMA00}. Projected exclusion plots for the CDMS \cite
{CDMS-PRO}, GENIUS \cite{Klac01} and GEM-I experiments are depicted too. The
large shaded area represents the theoretical prediction for allowed
spin-independent elastic WIMP-proton scattering cross section calculated in 
\cite{Ell00}.}
\end{center}
\end{figure}
\hspace{1.0in}

Hence, we can conclude that challenging scientific goal to touch $\approx $
0.01~eV neutrino mass domain, would be indeed feasible for the GEM project,
which realization seems to have practically no technical risk. To this
effect, the possibility to use already existing BOREXINO CTF as outer water
shield is very attractive. The GEM experiment will bring the outstanding
results for the 2$\beta $ decay studies (GEM-I and GEM-II stages) as well as
for the dark matter searches (GEM-I), which are of great interest and would
provide crucial tests of the many key problems and theoretical models of the
modern astroparticle physics.

\hspace{1.0in}

The authors express their gratitude to Y. Ramachers for fruitful discussions
and remarks and S.Yu.~Zdesenko for valuable suggestions concerning design of
the GEM set up.


\begin{thebibliography}{999}
\bibitem{Zub98}  K. Zuber, Phys. Rep. 305 (1998) 295.

\bibitem{Val82}  J. Schechter and J.W.F. Valle, Phys. Rev. D 25 (1982) 2951.

\bibitem{Moe94}  M.~Moe and P.~Vogel, Ann. Rev. Nucl. Part. Sci. 44 (1994)
247.

\bibitem{Tre95}  V.I.~Tretyak and Yu.G.~Zdesenko, At. Data Nucl. Data Tables
61 (1995) 43.

\bibitem{Fae98}  A.~Faessler and F.~Simkovic, J. Phys. G: Nucl. Part. Phys.
24 (1998) 2139.

\bibitem{Klap98}  H.V. Klapdor-Kleingrothaus, Int. J. Mod. Phys. A 13 (1998)
3953;

M. Hirsch and H.V. Klapdor-Kleingrothaus, Prog. Part. Nucl. Phys. 40 (1998)
323.

\bibitem{Suh98}  J.~Suhonen and O.~Civitarese, Phys. Rep. 300 (1998) 123.

\bibitem{Vog00}  P. Vogel, nucl-th/0005020 9 May 2000.

\bibitem{Kir99}  T.A. Kirsten, Rev. Mod. Phys. 71 (1999) 1213.

\bibitem{SK98}  Y. Fukuda et al. (Super-Kamiokande Collaboration), Phys.
Rev. Lett. 81 (1998) 1562; 82 (1999) 1810; 82 (1999) 2430.

\bibitem{LSND00}  E.D. Church (for the LSND Collaboration), Nucl. Phys. A
663\&664 (2000) 799; A. Aguilar et al., hep-ex/0104049 27 Apr 2001.

\bibitem{Klap00}  H.V. Klapdor-Kleingrothaus et al., Phys. Rev. D 63 (2001)
073005; hep-ph/0003219 v4 8 Oct 2000.

\bibitem{Bil99}  S.M. Bilenkij et al., Phys. Lett. B 465 (1999) 193.

\bibitem{Se82}  S.R. Elliot et al., Phys. Rev. C 46 (1992) 1535.

\bibitem{Mo100}  H. Ejiri et al., Phys. Rev. C 63 (2001) 065501.

\bibitem{Cd2000}  F.A.~Danevich et al., Phys. Rev. C 62 (2000) 045501.

\bibitem{Te130}  A. Alessandrello et al., Phys. Lett. B 486 (2000) 13.

\bibitem{Xe136}  R. Luescher et al., Phys. Lett. B 434 (1998) 407.

\bibitem{Ge76}  L. Baudis et al., Phys. Rev. Lett. 83 (1999) 41.

\bibitem{IGEX}  C.E. Aalseth et al., Phys. Rev. C 59 (1999) 2108;

D. Gonzalez et al., Nucl. Phys. B (Proc. Suppl.) 87 (2000) 278.

\bibitem{GEANT}  R. Brun et al., CERN Program Library Long Write-up W5013,
CERN, 1994.

\bibitem{Decay4}  O.A.~Ponkratenko et al., Phys. Atom. Nucl. 63 (2000) 1282.

\bibitem{Mat66}  E. der Mateosian and M. Goldhaber, Phys. Rev. 146 (1966)
810.

\bibitem{Fio67}  E. Fiorini et al., Phys. Lett. B 25 (1967) 607.

\bibitem{Fio70}  E. Fiorini et al., Lett. Nuovo Cimento vol. III, n. 5
(1970) 149.

\bibitem{Isotop}  A.A. Artyukhov et al., Phys. Atom. Nucl. 61 (1998) 1236;

A. Pokidychev, M. Pokidycheva, Nucl. Instrum. Meth. A 438 (1999) 7.

\bibitem{Moe91}  M.K. Moe, Phys. Rev. C 44 (1991) 931.

\bibitem{Miya91}  M. Miyajima et al., KEK Proc. 91-5 (1991) 19.

\bibitem{Miya96}  M. Miyajima et al., AIP Conf. Proc. 338 (1997) 253.

\bibitem{EXO}  M. Danilov et al., Phys. Lett. B 480 (2000) 12.

\bibitem{Ragh94}  R.S. Raghavan, Phys. Rev. Lett. 72 (1994) 1411.

\bibitem{Xe-CTF}  B. Caccianiga, M.G. Giammarchi, Astropart. Phys. 14 (2000)
15.

\bibitem{MOON}  H. Ejiri et al., Phys. Rev. Lett. 85 (2000) 2917.

\bibitem{Gd-160}  F.A.~Danevich et al., Nucl. Phys. B (Proc. Suppl.) 48
(1996) 235;

F.A.~Danevich et al., nucl-ex/0011020 24 Nov 2000; Nucl. Phys. A (2001) in
press.

\bibitem{Rag97}  R.S. Raghavan, Proc. 4th Int. Solar Neutrino Conf.,
Heidelberg, Germany, 8-11 April 1997. - Max-Planck-Institut fur Kernphysik,
Heidelberg, 1997, p. 248.

\bibitem{Zub00}  K. Zuber, Phys. Lett. B 485 (2000) 23.

\bibitem{Pig99}  F. Piquemal for the NEMO collab., Nucl. Phys. B (Proc.
Suppl) 77 (1999) 352.

\bibitem{Fio98}  E. Fiorini, Phys. Rep. 307 (1998) 309.

\bibitem{Nemo00}  NEMO Collaboration, hep-ex/0006031 26 June 2000.

\bibitem{Ger00}  G. Gervasio (for the CUORE collaboration), Nucl. Phys. A
663\&664 (2000) 873.

\bibitem{CAMEO}  G. Bellini et al., Phys. Lett. B 493 (2000) 216.

\bibitem{MAJOR}  Majorana project website: http://majorana.pnl.gov.

\bibitem{GENIUS-98}  H.V.~Klapdor-Kleingrothaus et al., J. Phys. G: Nucl.
Part. Phys. 24 (1998) 483.

\bibitem{Bau99}  L. Baudis et al., Nucl. Instrum. Meth. A 426 (1999) 425.

\bibitem{Pon98}  O.A. Ponkratenko et al., Proc. Int. Conf. on Dark Matter in
Astro and Particle Phys., Heidelberg, Germany, 20-25 July 1998, eds. H.V.
Klapdor-Kleingrothaus and L. Baudis, IOP, Bristol, Philadelphia, 1999, p.738.

\bibitem{Bel96}  G.~Bellini (for the BOREXINO Collaboration), Nucl. Phys. B
(Proc. Suppl.) 48 (1996) 363;

G. Alimonti et al., Nucl. Instrum. Meth. A 406 (1998) 411.

\bibitem{Vacuum}  R.H. Kropschot, Cryogenics 1 (1961) 171.

\bibitem{Gun97}  M.~Gunther et al., Phys. Rev. D 55 (1997) 54;

L.~Baudis et al., hep-ex/0012022 7 Dec 2000.

\bibitem{Jag93}  P.~Jagam and J.J.~Simpson, Nucl. Instrum. Meth. A 324
(1993) 389.

\bibitem{Heus00}  G. Heusser et al., Appl. Rad. and Isotopes 52 (2000) 691.

\bibitem{SNO}  A.B. McDonald, Nucl. Phys. B (Proc. Suppl.) 77 (1999) 43;

J. Boger et al., Nucl. Instrum. Meth. A 449 (2000) 172.

\bibitem{Mar92}  C.J.~Martoff and P.D.~Lewin, Comp. Phys. Comm. 72 (1992) 96.

\bibitem{Gri91}  {\it Physical Quantities: The Handbook}, eds.
I.S.~Grigoriev et al., Energoatomizdat, Moscow, 1991.

\bibitem{arpesella92}  C. Arpesella, Nucl. Phys. A 28 (1992) 420.

\bibitem{KamLand}  A. Suzuki, Nucl. Phys. B (Proc. Suppl.) 77 (1999) 171.

\bibitem{Jun96}  G. Jungmann et al., Phys. Rep. 267 (1996) 195.

\bibitem{Ram99}  Y. Ramachers, astro-ph/9911260 15 Nov 1999.

\bibitem{Bau00a}  L. Baudis and H.V.~Klapdor-Kleingrothaus, astro-ph/0003434
29 Mar 2000.

\bibitem{Kla01}  H.V.~Klapdor-Kleingrothaus, hep-ph/0102319 26 Feb 2001.

\bibitem{Klap01}  H.V.~Klapdor-Kleingrothaus et al., hep-ph/0103082 8 Mar
2001.

\bibitem{Doi83}  M. Doi et al., Prog. Theor. Phys. Suppl. 69 (1983) 602;
Prog. Theor. Phys. 89 (1993) 139.

\bibitem{Kla97c}  H.V.~Klapdor-Kleingrothaus, M. Hirsch, Z. Phys. A 359
(1997) 361.

\bibitem{Riz96}  T. Rizzo, SLAC-PUB-7365, hep-ph/9612440 20 Dec 1996; M.
Cvetic and S. Godfrey, hep-ph/9504216 4 Apr 1995; S. Godfrey et al.,
hep-ph/9704291 10 Apr 1997.

\bibitem{H196}  H1 Collab., S. Aida et al., Phys. Lett. B 369 (1996) 173.

\bibitem{Hir96}  M. Hirsch et al., Phys. Lett. B 378 (1996) 17; Phys. Rev. D
54 (1996) 4207.

\bibitem{MPI99}  H.V.~Klapdor-Kleingrothaus et al., MPI-Report
MPI-H-V26-1999, Heidelberg, 1999.

\bibitem{Pan97}  N. Cabibbo et al., Phys. Lett. B 139 (1984) 459;

O. Panella et al., Phys. Rev. D 56 (1997) 5766.

\bibitem{Pan99}  O. Panella et al., Phys. Rev. D 62 (2000) 015013.

\bibitem{Moh86}  R. Mohapatra, Phys. Rev. D 34 (1986) 3457.

\bibitem{Hir95}  M. Hirsch et al., Phys. Rev. Lett. 75 (1995) 17; Phys. Rev.
D 53 (1996) 1329; Phys. Lett. B 372 (1996) 181; Phys. Lett. B 459 (1999) 450.

\bibitem{Fae97}  A. Faessler et al., Phys. Rev. Lett. 78 (1997) 183; Phys.
Rev. D 58 (1998) 055004; Phys. Rev. D 58 (1998) 115004.

\bibitem{Wod99}  A. Wodecki et al., Phys. Rev. D 60 (1999) 115007.

\bibitem{Hir97}  M. Hirsch et al., Phys. Lett. B 398 (1997) 311; 403 (1997)
291; Phys. Rev. D 57 (1998) 2020.

\bibitem{Hir98b}  M. Hirsch et al., Phys. Rev. D 57 (1998) 1947.

\bibitem{Bha99}  G. Bhattacharyya et al., Phys. Lett. B 463 (1999) 77.

\bibitem{Vis99}  F. Vissani, JHEP 9906 (1999) 022.

\bibitem{Cza99}  M. Czakon et al., Acta Phys. Pol. B 30 (1999) 3121.

\bibitem{Cza00}  M. Czakon et al., hep-ph/0010077 9 Oct 2000.

\bibitem{Czak00}  M. Czakon et al., Acta Phys. Pol. B 31 (2000) 1365.

\bibitem{Bil01}  S.M. Bilenky et al., hep-ph/0102265 21 Feb 2001;
hep-ph/0104218 23 Apr 2001.

\bibitem{Klab01}  H.V.~Klapdor-Kleingrothaus, hep-ph/0102276 22 Feb 2001;
hep-ph/0103074 7 Mar 2001.

\bibitem{Klac01}  H.V.~Klapdor-Kleingrothaus, B. Majorovits, hep-ph/0103079
7 Mar 2001.

\bibitem{Ell00}  J. Ellis et al., Phys. Lett. B 481 (2000) 304.

\bibitem{Bed01}  V. A. Bednyakov, H.V.~Klapdor-Kleingrothaus, Phys. Rev. D
63 (2001) 095005.

\bibitem{H-M99}  L. Baudis et al., Phys. Rev. D 59 (1999) 022001.

\bibitem{DAMA98}  R. Bernabei et al., Nucl. Phys. B (Proc. Suppl.) 70 (1998)
79; Phys. Lett. B 389 (1996) 757.

\bibitem{CDMS00}  R. Abusaidi et al., Phys. Rev. Lett. 84 (2000) 5699.

\bibitem{DAMA00}  R. Bernabei et al., Phys. Lett. B 480 (2000) 23.

\bibitem{CDMS-PRO}  R. Abusaidi et al., Nucl. Inst. Meth. A 444 (2000) 345.

\bibitem{Ceb01}  S. Cebrian et al., Astrop. Phys. 14 (2001) 339.

\hspace{1.0in}

\hspace{1.0in}
\end{thebibliography}
\end{document}